\documentclass[11pt]{article}
\pdfoutput=1
\usepackage{jheppub}
\usepackage{graphicx,float,flowchart}
\usepackage{amssymb}
\usepackage{subcaption}
\usepackage{amsmath}
\usepackage{slashed}
\usepackage{hyperref}
\usepackage{caption}
\usepackage{xcolor}
\usepackage{dsfont}
\usepackage{verbatim}
\usepackage{setspace}
\usepackage{slashed}
\usepackage{xcolor}
\usepackage{braket}

\newcommand\beq{\begin{equation}}
\newcommand\eeq{\end{equation}}

\newcommand{\Tr}{{\rm Tr\,}}

\def\shrug{\texttt{\raisebox{0.75em}{\char`\_}\char`\\\char`\_\kern-0.5ex(\kern-0.25ex\raisebox{0.25ex}{\rotatebox{45}{\raisebox{-.75ex}"\kern-1.5ex\rotatebox{-90})}}\kern-0.5ex)\kern-0.5ex\char`\_/\raisebox{0.75em}{\char`\_}}}

\title{Some Information Theoretic Aspects of De-Sitter Holography}
\author{Hao Geng}
\affiliation{Department of Physics, University of Washington, Seattle, WA, 98195-1560, USA}
\emailAdd{hg666@uw.edu}
\preprint{\today}
\abstract{Built on our observation that entangling surfaces of the boundary field theory are great co-dimension one spheres in the context of DS/dS correspondence, we study some information theoretic quantities of the field theory dual intensively using holographic proposals. We will focus on entanglement entropy (EE), entanglement of purification (EoP) and complexity. Several fundamental observations and analysis are provided. For EE, we focus on its scaling behavior, which indicates the nature of the relevant degrees of freedom. Moreover, we find that EE provides us with important information of the energy spectrum in pure dS and it also leads us to the speculation that the field theory dual is chaotic or non-integrable. For EoP, an interesting phenomenon we call "Entanglement Wedge Cross Section (EWCS) Jump" is observed according to which we propose two puzzles regarding EoP and EE in the context of dS holography. For complexity, we find that the Complexity=Volume proposal does not provide a well-defined way to compute complexity for pure dS. However, it does provide a well-defined way to compute complexity in the $T\bar{T}+\Lambda_{2}$ deformed case. At the end, we will use the surface/state correspondence to resolve all the puzzles and hence reach a consistent information theoretic picture of dS holography. Moreover, we will provide evidence for our former proposal that the $T\bar{T}+\cdots$ deformations are operating quantum circuits and study the non-locality of the field theory algebra suggested by the surface/state correspondence.}
\begin{document}
\maketitle
\pagebreak
\section{Introduction}
In this section we will discuss some background ideas, set up our notation and outline the structure of this paper. 

The basic motivation for the DS/dS correspondence is from the observation of \cite{Karch:2003em} that de-Sitter (dS) space can be put into the form of a Randall-Sundrum construction \cite{Randall:1999vf} as two asymptotically chopped anti-de-Sitter (AdS) spaces glued together along a UV brane which localizes a graviton through the Randall-Sundrum mechanism. Based on this observation, it is proposed in \cite{Alishahiha:2004md} that the holographic dual description of $dS_{D}$ quantum gravity is a system of two conformal field theories (CFT) living on the UV brane coupled to each other by the localized graviton as a D-1-dimension gravitational system.

This is supported by the observation that the metric of $dS_{d}$ sliced global $(A)DS_{d+1}$ is
\begin{equation}
\label{metric}
    ds^{2}_{(A)DS_{d+1}}=d\omega^{2}+\sin(h)^{2}(\frac{\omega}{L})(-d\tau^{2}+L^{2}\cosh^{2}(\frac{\tau}{L})d\Omega_{d-1}^{2})
\end{equation}
where $\omega$ is the radial coordinate in AdS case which is an energy scale in the AdS/CFT correspondence \cite{Aharony:1999ti} and in the context of that $\omega\rightarrow\infty$ is the ultra-violet (UV) and $\omega=0$ is the infrared (IR). AdS/CFT defines quantum gravity in asymptotically AdS spaces. From the metric we just wrote down we see that, from the holographic point of view, de-Sitter quantum gravity and anti-de-Sitter quantum gravity should have the same near horizon (IR) physics which is known to be the IR of the CFT living on the asymptotic boundary of AdS. However, in dS case the UV is chopped and we have two identical copies of space glued to each other through the UV brane so naturally we have two UV-cutoff CFTs living on the brane coupled to each other through the residual graviton localized by the brane.

Even though this picture is beautiful, we almost know nothing about how the DS/dS correspondence operates. For example we do not know much about the dictionary between bulk fields and boundary operators and the protocol to compute field theory correlators\footnote{I thank Eva Silverstein for pointing out that some very recent developments addressing these questions are made in \cite{Lewkowycz:2019xse} and references therein.}. However, some recent primitive efforts have been made from the information theoretic point of view focusing on entanglement entropy \cite{Dong:2018cuv,Geng:2019bnn} and the emergence of pure dS space \cite{Gorbenko:2018oov}. Several novel properties and behaviors of holographic entanglement entropy in the context of the DS/dS correspondence are identified in these two works. For example, it is observed in \cite{Geng:2019bnn} that the symmetry of dS space provides an infinite amount of RT surfaces \cite{Ryu:2006bv} equally calculating the entanglement entropy of the CFT-system. More than this, the null-energy condition (NEC) says that one of them, which is supposed to calculate CFT entropy, is always greater than another one among them, which calculates the Gibbons-Hawking entropy for the static patch, when we turn on matter deformations in the bulk. Beyond this, as we will discuss in Sec.\ref{sec:matterdeform},  when $D=3$ and we turn on massless scalar deformation the surface calculating the static patch Gibbons-Hawking entropy has the smallest area among all the minimal area surfaces. However, in general dimension and for arbitrary deformation there is nothing we are able to say, at least for now.

In this paper, we will follow-up on these observations and the proposal that $T\bar{T}+\Lambda_{2}$ deformation of the UV-cutoff CFT generates the bulk of dS space with deformation parameter dual to the bulk radial coordinate in \cite{Gorbenko:2018oov, McGough:2016lol} to study more information theoretic quantities- entanglement of purification (EoP) and holographic complexity \cite{Nguyen:2017yqw,Takayanagi:2017knl, Alishahiha:2015rta}. As a warm-up and a review, we start from our familiar case - entanglement entropy (EE) - studying its scaling behavior with the subsystem size before and after we turn on the $T\bar{T}+\Lambda_{2}$ deformation and then move on to the other two. A novel behavior of EoP, we call ``entanglement wedge cross section (EWCS) jump'' is discovered and according to it two puzzles regarding EoP and EE are proposed and we find the complexity=volume (CV) proposal in \cite{Alishahiha:2004md} might have to be refined in a non-trivial way as a general holographic dictionary. All these puzzles can be resolved using the surface/state correspondence proposed in \cite{Miyaji:2015yva}. As a bonus, this analysis supports our observation in \cite{Geng:2019yxo} that the $T\bar{T}+\cdots$ deformations are operating a quantum circuit which prepares quantum states.

The structure of this paper is that we will start from a review of the basic concept, definition and proposed holographic dual of the information theoretic quantity we study in each section and then move on to study its behavior in the context of the DS/dS correspondence and end up with a discussion. We try to write each section to be independent of the others for readers with different interests. We hope to clarify our observation and provide our primitive discussion in an elegant way so we'll avoid complicated computations in higher dimensions and always focus on $dS_{3}$. Finally, we will briefly review the surface/state correspondence and use it to resolve the puzzles we encountered and provide several evidence, relations and extensions of the previous works \cite{Geng:2019yxo,Lewkowycz:2019xse}.

In our dS computations for \eqref{metric} we will take $\tau=0$ and $d=2$ so $d\Omega_{d-1}=d\phi$ where $\phi\in[0,2\pi]$ and $\frac{\omega}{L}\in[0,\pi]$. We emphasis that $\frac{\omega}{L}=0,\pi$ are two horizons identical to that in AdS and $\frac{\omega}{L}=\pi/2$ is the central slice (UV brane) where the two CFTs live.
\section{Scaling behavior of Entanglement Entropy}
The entanglement entropy is a primitive information theoretic quantity assessing how tightly are two systems entangled to each other. The entanglement entropy of a subsystem $\mathcal{A}$ and its complement is simply defined as
\begin{equation}
    S_{\mathcal{A}}=-\Tr\rho_{\mathcal{A}}\log\rho_{\mathcal{A}}
\end{equation}
where $\rho_{\mathcal{A}}$ is the reduced density matrix of the subsystem $\mathcal{A}$. It was extensively studied from several different points of view in the last decade. Movitated by the Bekenstein-Hawking entropy formula \cite{Hawking:1974sw,Bekenstein:1973ur}, it is proposed in \cite{Ryu:2006bv} that for a field theory dual of asymptotic AdS gravity the entanglement entropy of a subsystem $\mathcal{A}$ and its complement is given by a quarter area of the bulk minimal surface anchored on the boundary $\partial\mathcal{A}$ and homologous to $\mathcal{A}$ in Plank unit. However, it is believed that information theoretic quantities capture some general properties of the system we are studying and it is well believed that proposals like \cite{Ryu:2006bv} are generally true in quantum gravity. In this section, we use the RT proposal to discuss the scaling property of entanglement entropy, in the context of dS holography, with the subsystem size before and after we lift the field theory dual into the bulk through $T\bar{T}+\Lambda_{2}$ deformation \cite{Gorbenko:2018oov}.
\subsection{Pure dS}
A novel behavior of RT surfaces for pure dS holography uncovered in \cite{Geng:2019bnn} is that for a (connected) subsystem $\mathcal{B}$, on the field theory side, whose size is exactly half of the whole central slice, there are an infinite number of entangling surfaces which calculate exactly the same amount of EE\footnote{Other interesting recent studies of dS entanglement entropy are \cite{Arias:2019zug,Arias:2019pzy}.}. The existence of these entangling surfaces is supported by the symmetry of dS space which is a very important intrinsic property of its geometry. In this case, the spatial geometry of \eqref{metric} can be visualized as a sphere where the central equator is the place the codimension-one field theory system living. Hence for a subsystem $\mathcal{A}$ its RT surface is the bulk great circle anchored on its boundary and so if $\mathcal{A}$ is exactly half of the equator it has an infinite number of such surfaces. These surfaces have areas exactly equal to the volume (length in $dS_{3}$ case) of $\mathcal{A}$. As a result, the EE is scaling with the size of $\mathcal{A}$ as its volume.
\subsection{Cutoff dS}
If we move the field theory system into the bulk using the $T\bar{T}+\Lambda_{2}$ deformation, then for subsystem $\mathcal{A}$ now living on the cutoff slice $\omega=\omega_{c}$ there is always one RT surface no matter how large $\mathcal{A}$ is. Furthermore, we can still use our observation that minimal surfaces are great circles to find the RT surface for any $\mathcal{A}$ (see Fig.\ref{pic:RTcutoff}). More details and calculations in higher dimensions could be found in \cite{Grieninger:2019zts}. In our $dS_{3}$ case, assuming the angular size of $\mathcal{A}$ is $\Delta{\theta}$, we find that the entanglement entropy is
\begin{equation}
    S_{\mathcal{A}}=\frac{l_{\mathcal{A}}}{4G_{N}}
\end{equation}
where $G_{N}$ is Newton constant,
\begin{equation}
\begin{split}
    l_{\mathcal{A}}&=L\int\sqrt{d(\frac{\omega}{L})^{2}+\sin^{2}(\frac{\omega}{L})d\phi^{2}}\\
    &=\frac{2L}{\sin(\frac{\omega_{*}}{L})}\int_{\frac{\pi-\Delta\theta}{2}}^{\frac{\pi}{2}}d\phi\frac{1}{1+\sin^{2}\phi\cot^{2}\frac{\omega_{*}}{L}}\\
    &=2L\cot^{-1}(\frac{\cot\frac{\Delta\theta}{2}}{\sin\frac{\omega_{*}}{L}})
    \end{split}
\end{equation}
and $\omega_{*}$ is the minimal bulk radial coordinate approached by the RT surface determined by
\begin{equation}
    \cos(\frac{\Delta\theta}{2})=\frac{\cot\frac{\omega_{c}}{L}}{\cot\frac{\omega_{*}}{L}}.
\end{equation}
In summary, the entanglement entropy for the field theory subsystem $\mathcal{A}$ with angular size $\Delta\theta$ living on the cutoff slice $\omega_{c}$ is
\begin{equation}
    S(\Delta\theta,\omega_{c})=\frac{L\cot^{-1}(\sqrt{\cot^{2}\frac{\Delta\theta}{2}+\frac{\cot^{2}\frac{\omega_{c}}{L}}{\sin^{2}\frac{\Delta\theta}{2}}})}{2G_{N}}.\label{eq:entropysize}
\end{equation}
\subsection{Discussion}
In this subsection, we will discuss lessons we learn from the behavior of EE for the field theory dual of dS gravity.   
\subsubsection{Scaling Behavior-Relevant Degrees of Freedom}
Entanglement entropy is supposed to capture properties of the interaction among the underlying degrees of freedom of the theory. In our context, dS space has a horizon and so its field theory dual would be of finite temperature. Together with the realization that the field theory dual is large N (because the gravity side is a spacetime described by a solution of Einstein's equation \cite{Harlow:2018fse}), we have the following picture that if EE of the subsystem is scaling as its volume then the interaction among the underlying degrees of freedom is weak and the system is a thermal gas so the effective degrees of freedom are the same as the underlying deconfined or local ones (say fundamental particles). But if EE is scaling as the subsystem's area (its co-dimension one volume measure) or even lower then the interaction should be strong and hence the effective degrees of freedom of the system should be confined or nonlocal (say strings or glueballs) and hence causes the effective dimensional reduction from volume behavior to area and/or lower. More concretely, this picture can be understood by the following behavior of EE of a subsystem in a large N field theory with finite temperature\footnote{I thank Andreas Karch for providing this illustrative example and Kristan Jensen for some critical comments.}
\begin{equation}
    S=\begin{cases}
       \alpha N^{2}V+\gamma N^{2}A+\cdots,\text{  deconfined}
       \\\alpha' V+\gamma' N^{2}A+\cdots,\text{  confined}
    \end{cases}
\end{equation}
where V and A are volume and boundary area of the subsystem and $\alpha$, $\alpha'$, $\gamma$ and $\gamma'$ are $O(1)$ constants.

\begin{figure}[H]
    \centering
   \includegraphics[width=10.1cm]{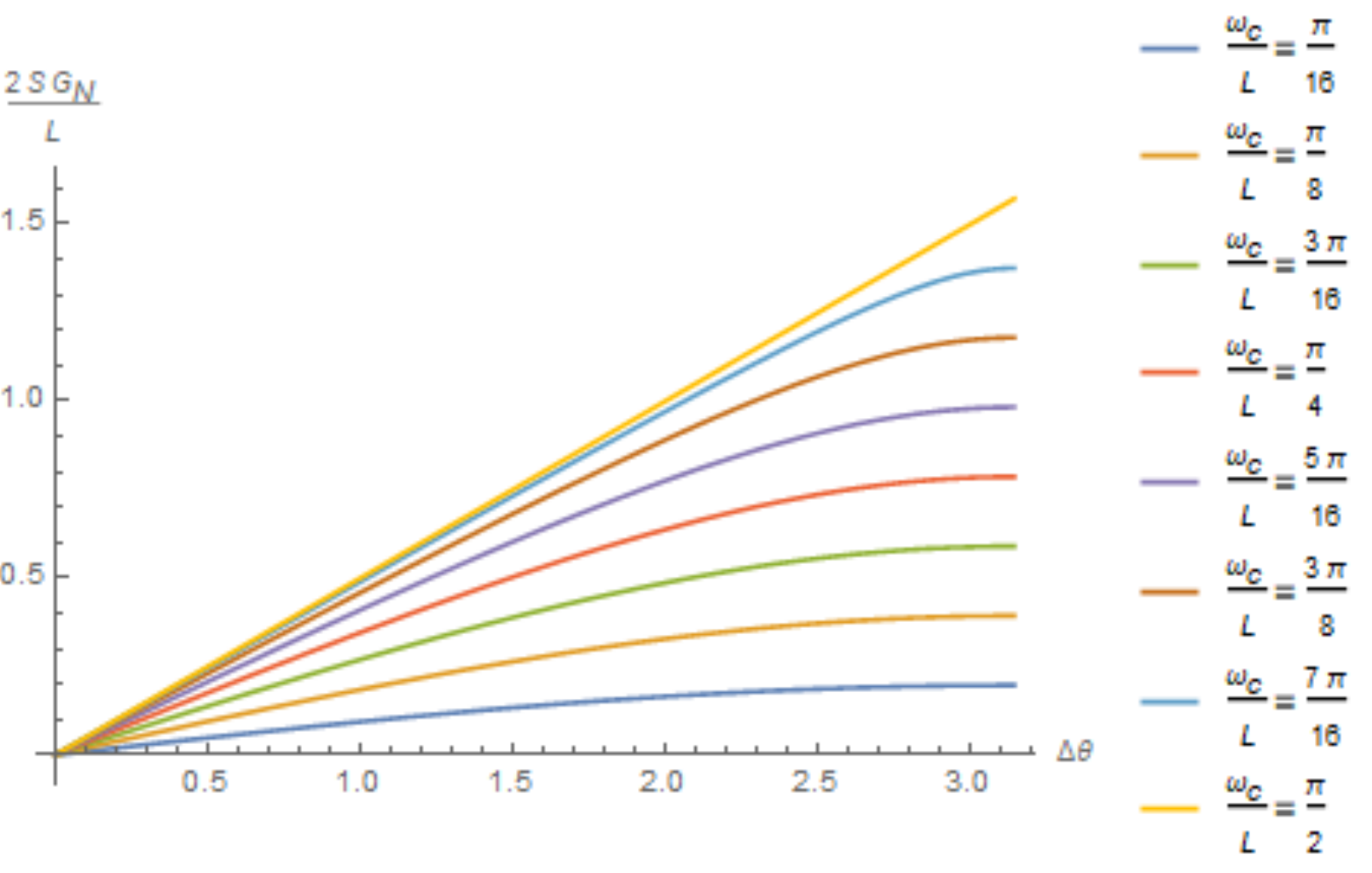}
        \caption{The Scaling Behavior of EE}\label{pic:ScalingEE}
\end{figure}
The scaling behavior of EE in our context is summarized in Fig.\ref{pic:ScalingEE}. As we can see that if we go deeper and deeper into the bulk its dependence on the subsystem size is weaker and weaker and so the interaction among the field theory's underlying degrees of freedom is stronger and stronger. This picture is in accord with the general proposal in gauge/gravity duality that the field theory controlled by a strongly interacting IR fixed point should be dual to a gravitational theory near horizon. Furthermore, it is in accord with the observation that the $T\bar{T}$ type deformations do not produce local quantum field theory \cite{Cardy:2019qao,Lewkowycz:2019xse}.

It is interesting to realize that $T\bar{T}+\Lambda_{2}$ is an irrelevant deformation which does not modify the IR physics of field theory. Therefore, it provides us a tool to probe some IR physics of the strongly coupled field theory from the gravity side.
\subsubsection{Energy Spectrum for Pure dS}
\paragraph{Small Subsystems}
For pure dS, without cutoff, if we restrict our discussion to subsystems with size smaller than half of the whole central slice, then the RT surfaces would just be the subsystems themselves. As a result, the entanglement wedges (for a quick definition see \ref{sec:EW}) are empty. This means that the density matrices of these subsystems are proportional to identity because for each of these subsystems the density matrix commutes with all local operators inside it (see \cite{Faulkner:2017vdd} for more of this point of view and the entanglement wedge). Hence the states of subsystems with size smaller than half of the whole are maximally mixed (this is confirmed in \cite{Lewkowycz:2019xse} for $DS_{3}$ from an explicit field theory calculation). Together with our discussions in the previous subsection that these subsystems should be thermal as they are inside the $dS_{d}$ static patch horizon, we claim that the energy spectrum of them is highly degenerate to zero.
\paragraph{The Whole System} \label{sec:ergodic}The CFT system living on the central slice is generated by the $T\bar{T}+\Lambda_{2}$ deformation such that the deformation is large enough that the system is chaotic. This is because the central slice is the only place that the field theory system living on it has entanglement entropy scaling linearly with the subsystem size i.e. the phase space is wholly exploited with equal probability or the ergodicity is satisfied \cite{DAlessio:2016rwt}. As a result, the energy spectrum of this system should satisfy the Wigner surmise \cite{DAlessio:2016rwt} and the field theory system is likely not integrable. This leads us to question the integrability of the $T\bar{T}+\Lambda_{2}$ deformed field theory dual. Understanding this point better will be helpful for future studies of $dS_{3}$ quantum gravity and the $DS_{3}/dS_{2}$ holography, for example studying the partition function. 
\section{Entanglement of Purification}
In this section the bipartite system  we consider consists of two disjoint intervals A and B with equal size and antipodally on the slice they live. We restrict our attention to subsystems with size smaller than half of the whole slice they live. We denote the RT surface of $\text{A}\cup \text{B}$ as $\Sigma$ and the entanglement wedge (for definition see \ref{sec:EW}) on the time-slice $\tau=0$ as EW. For a bipartite system $\mathcal{H}_{\text{A}}\otimes\mathcal{H}_{\text{B}}$ with density matrix $\rho_{\text{AB}}$ and purifier A$'$B$'$ in the purified state $\ket{\psi}$, the entanglement of purification is defined as \cite{2002JMP....43.4286T}:
\begin{equation}
    EoP(\text{A}:\text{B})=\underset{\psi,\text{A'}}{\min}\text{ } S_{AA'}\label{eq:EoPdef}
\end{equation}
where the minimization is over all purifiers and bipartitions of each purifier into A$'$ and B$'$ and $S_{\text{AA}'}$ is the entanglement entropy of AA$'$ and its complement BB$'$ in the whole pure system. Conceptually, EoP measures the amount of correlations in a quantum state of the system A$\cup$B for both classical and quantum correlations. Therefore, if EoP is zero then there is completely no correlation between A and B and so they do not know the existence of each other and are not able to communicate.

\begin{figure}[H]
    \centering
    \includegraphics[width=8cm]{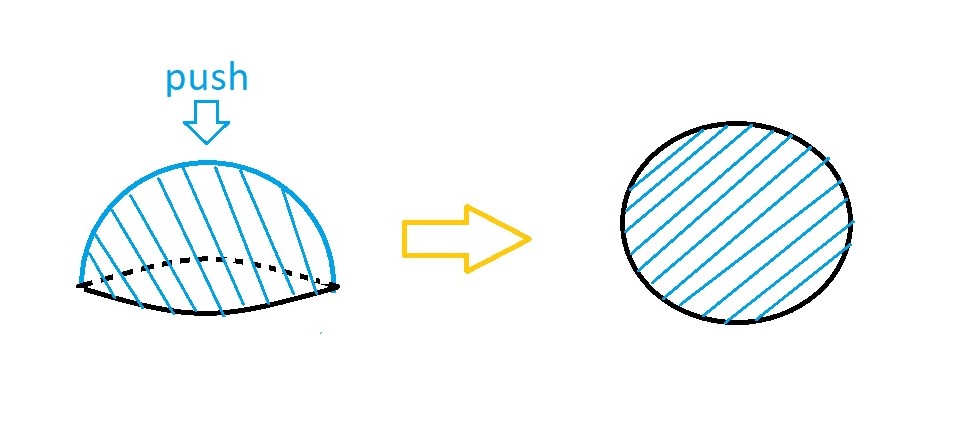}
    \caption{Push down the bulk hemisphere onto the plane. The equator (central slice for the whole sphere or dS$_{3}$ $\tau=0$ time slice) on the left corresponds to the outer circle on the right.}
      \label{pic:push down}
\end{figure}

It has been proposed independently in \cite{Takayanagi:2017knl,Nguyen:2017yqw} that EoP of this subsystem is calculated holographically by
\begin{equation}
\label{eq:EoP}
    EoP(\text{A}:\text{B})=\frac{Area(\text{X})}{4G_{N}}
\end{equation}
where $\text{X}$ is the neck of EW (see Fig.\ref{pic:EoP}) or entanglement wedge cross section (EWCS)\footnote{Other proposals for the field theory duals of this quantity includes reflected entropy \cite{Dutta:2019gen}, logarithmic negativity \cite{Kudler-Flam:2018qjo} and odd entanglement entropy \cite{Tamaoka:2018ned}.} In this section, we will use this proposal to study EoP of the field theory dual to pure dS and cutoff dS. An interesting phenomenon we call ``EWCS Jump" is identified along the way. ``Jump" contains two meanings here. On the one hand, for pure dS it says that the spectrum of EoP is discrete (which actually only contains two elements) and highly degenerate and so its value will jump between these two elements. On the other hand, if we lift the field theory up into the bulk by putting a Dirichlet wall, which corresponds to $T\bar{T}+\Lambda_{2}$ deformed field theory away from the UV cutoff, the spectrum of EoP will suddenly become continuous with an isolated point and much less degenerate. We claim that this novel behavior of EWCS motivates some puzzles in dS holography if we treat \ref{eq:EoP} seriously. For the sake of convenience of displaying figures, we will push down the bulk (hemi-)sphere onto the plane in the following two subsections (see Fig.\ref{pic:push down}). For the sake convenience of terminology we will not distinguish EoP and EWCS until we present the puzzles at the end this section.
\subsection{Pure dS- Discrete and Finite Spectrum of EoP}
\label{sec:EW}
\begin{figure}
    \centering
    \includegraphics[width=5cm]{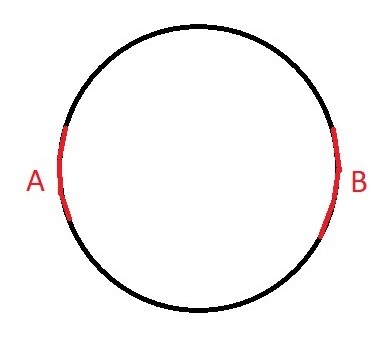}\quad 
    \includegraphics[width=5cm]{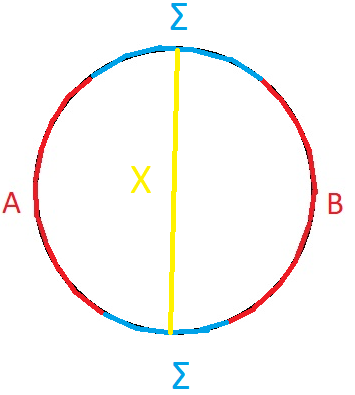}
        \caption{\textbf{Left}: If A and B are small then the RT surface for $\text{A}\cup\text{B}$ is $\text{A}\cup\text{B}$ so the entanglement wedge in bulk time slice is empty. There is no neck inside entanglement wedge meaning that EoP is zero. \textbf{Right}: If A and B are large enough then the RT surface of $\text{A}\cup\text{B}$ is its complement $\Sigma$ on the circle so the entanglement wedge is the bulk hemisphere. And the neck is a semi-great circle X stretching through the bulk.}\label{pic:EoP}
\end{figure}

In \cite{Geng:2019bnn} we observed that the RT surfaces of the field theory dual living on the central slice in the context of DS/dS are parts of great circles. As a result, the RT surface $\Sigma$ of $\text{A}\cup\text{B}$ is generally on the central slice having two disjoint components (see Fig.\ref{pic:EoP} where the outer circle is the equator of the sphere before we push the bulk down onto the plane in Fig.\ref{pic:push down}). The entanglement wedge EW of the subsystem A$\cup$B is defined as the bulk region enclosed by A$\cup$B and its RT surface $\Sigma$. If A and B are not large enough the RT surface $\Sigma$ is just A$\cup$B and so the region enclosed by $\Sigma$ and A$\cup$B, i.e. the EW, is empty (see Left of Fig.\ref{pic:EoP}). Then in this case, there is no neck of EW so $EoP(\text{A}:\text{B})$ is zero. However, on the hand, if A and B are large enough ($A\cup B$ is larger than or equal to half of the whole system) then the RT surface $\Sigma$ is the complement of A$\cup$B on the central slice which tells us that the EW is the whole bulk hemisphere (see Right of Fig.\ref{pic:EoP}). Hence, in this case, the neck of the EW is just any diameter of the circle stretching between the two components of $\Sigma$ which corresponds to a half great circle perpendicular to the equator in the spherical geometry. 
Using the geometric set up $\eqref{metric}$ and holographic proposal \eqref{eq:EoP}, we get the spectrum of EoP for subsystems smaller than half of the whole and put antipodally on the slice they live
\begin{equation}
    EoP(\text{A}:\text{B})=\begin{cases}
       0,\text{  if $A\cup B$ is smaller than half of the whole central slice }
       \\\frac{\pi L}{4 G_{N}},\text{   if $A\cup B$ is bigger than half of the whole central slice}
    \end{cases}.
\end{equation}

\subsection{Cutoff dS- Gapped and Continuous Spectrum of EoP}
\begin{figure}
    \centering
    \includegraphics[width=12cm]{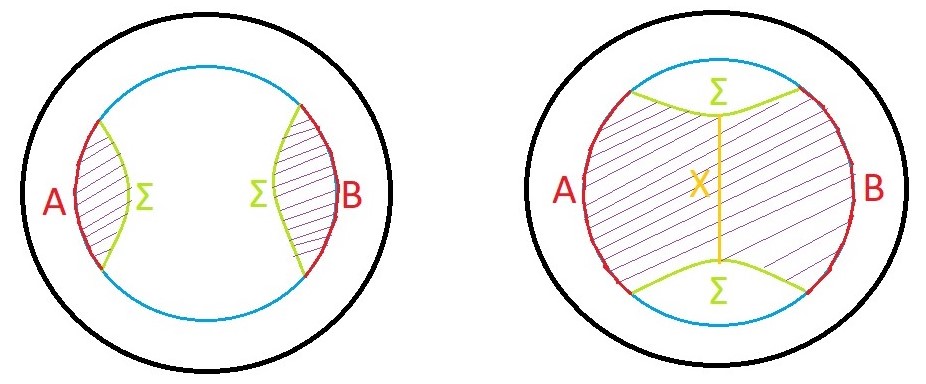}
        \caption{The cutoff slice is in blue and part of it- subsystems A and B in red. The shaded region is the EW. \textbf{Left}: If A and B are small and not close then the RT surface for $\text{A}\cup\text{B}$ is union of RT surfaces for A and B so the entanglement wedge is disconnected. Hence there is no neck inside the entanglement wedge meaning that EoP is zero. \textbf{Right}: If A and B are large and close enough then the RT surface of $\text{A}\cup\text{B}$ is different from the union of RT surfaces of A and B and the entanglement wedge is connected. The neck is part of a great circle X stretching through the EW.}\label{pic:intobulk}
\end{figure}

When we lift the field theory into the bulk, the spectrum and behavior of EoP is more like that in normal AdS where the neck of the entanglement wedge is unique \cite{Takayanagi:2017knl}. This can be seen from Fig.\ref{pic:intobulk}.
Let the Dirichlet wall to be put radially at $\omega_{c}=\alpha\pi L$ where $0<\alpha<\frac{1}{2}$. Then we see that the EoP has a  gapped and continuous spectrum
\begin{equation}
    \label{EoPconti}
    \text{EoP}\in\{0\}\cup [\delta_{c},\frac{L\alpha\pi}{2G_{N}}]
\end{equation}
where $\delta_{c}$ depends on $\omega_{c}$. EoP is zero when the two subsystems are not close or large enough so that the entanglement wedge is not connected even though it is nonempty. 
\subsection{Discussion}
In this section, we have provided two essential observations of the field theory dual of dS quantum gravity from a quantum information theoretic point of view. The similarity between the behaviors of EoP in pure AdS and cutoff dS should be understood as emphasizing the observation that quantum gravity in dS and AdS have the same near horizon (IR) description \cite{Alishahiha:2004md}. Another observation a novel behavior which we call ``EWCS Jump''.
\subsection{Some Puzzles}
In this subsection, we provide two puzzles regarding EoP jump in the case of pure dS.
\subsubsection{Puzzle One and a Possibly Refined Proposal of Holographic EoP}
In this section we have seen novel behaviors of the proposed EoP from holography. However, there is a puzzle for this proposal. As we know from its definition EoP should satisfy the following inequality \cite{Takayanagi:2017knl,Nguyen:2017yqw}
\begin{equation}
    EoP(\text{A}:\text{B})\leq \min{(S_{\text{A}},S_{\text{B}})}.\label{eq:ineq}
\end{equation}

However, for pure dS, in the case of the right on Fig.\ref{pic:EoP} $EoP$(A:B) is a quarter of the EW neck which is half as large as the whole system. Therefore $EoP$(A:B) is larger than a quarter of the size of either A or B and hence is bigger than the entanglement entropy of each of them. This violates the inequality \eqref{eq:ineq}. A possible refinement of the proposal \eqref{eq:EoP} is as follows
\begin{equation}
    EoP(\text{A}:\text{B})=\begin{cases}
       0,\text{  if EW($A\cup B$) is empty }
       \\\min{(\frac{Area(X)}{4 G_{N}},S_{A},S_{B})},\text{   if EW($A\cup B$) is not empty}\label{eq:refined}
       \end{cases}.
\end{equation}
\subsubsection{Puzzle Two- What Entropy Are We Calculating?}
In the DS/dS correspondence, we have two copies of CFT living on the central slice coupled to each other by a residual graviton. The RT surface which is the whole central slice is interpreted to calculate the entanglement entropy of the two CFTs or tracing out one CFT as discussed in \cite{Dong:2018cuv}. From this point of view, it is not hard to see that the EE is scaling as subsystem volume if the subsystem belongs to a single CFT even without the requirements of finite temperature and large N. Because then we always trace out its shadow on another CFT and they are locally entangled to each other by the residual graviton which is much stronger than the entanglement with other parts. However, if we take this interpretation for $A\cup B$ in Fig.\ref{pic:EoP} the same story is going on and hence we should always take $A\cup B$ as the entangling surface for $A\cup B$ no matter of what size it is. As a result, the EW is always empty and hence $EoP(A:B)$ is always zero. This definitely circumvents the puzzle one but there is no strong support that this is the right interpretation and it violates the Ryu-Takayanagi proposal \cite{Ryu:2006bv}. Hence the puzzle is that what entropy are we calculating if we use different RT surfaces discovered in \cite{Geng:2019bnn}.
\section{Complexity}
In this section, we will study holographic complexity using the complexity=volume (CV) proposal from \cite{Alishahiha:2015rta} which states that, in the holographic set-up, for a field theory subsystem $\mathcal{A}$ its complexity is proportional to the bulk volume of the region enclosed by $\mathcal{A}$ and its RT surface. Complexity is a quantity measuring how difficult it is to reach a state from a given state by fundamental unitary transformations. In has been proposed to be able to detect some properties of black holes in \cite{Susskind:2014rva} and been extensively studied these years (for some initial studies see references in \cite{Susskind:2014rva}). Besides its proposed definition, the physical meaning of it from the holographic point of view is still not clear. For example, what is it exactly dual on the field theory side? We'll not try to answer these questions or even make any comments on them (a recent realization that the CV proposal indeed calculates the complexity defined in quantum physics is \cite{Geng:2019yxo}). However, we would like to study if the CV proposal for computing this quantity is well-defined from the point of view of dS holography.

\subsection{Pure dS}
\begin{figure}[H]
    \centering
    \includegraphics[width=8cm]{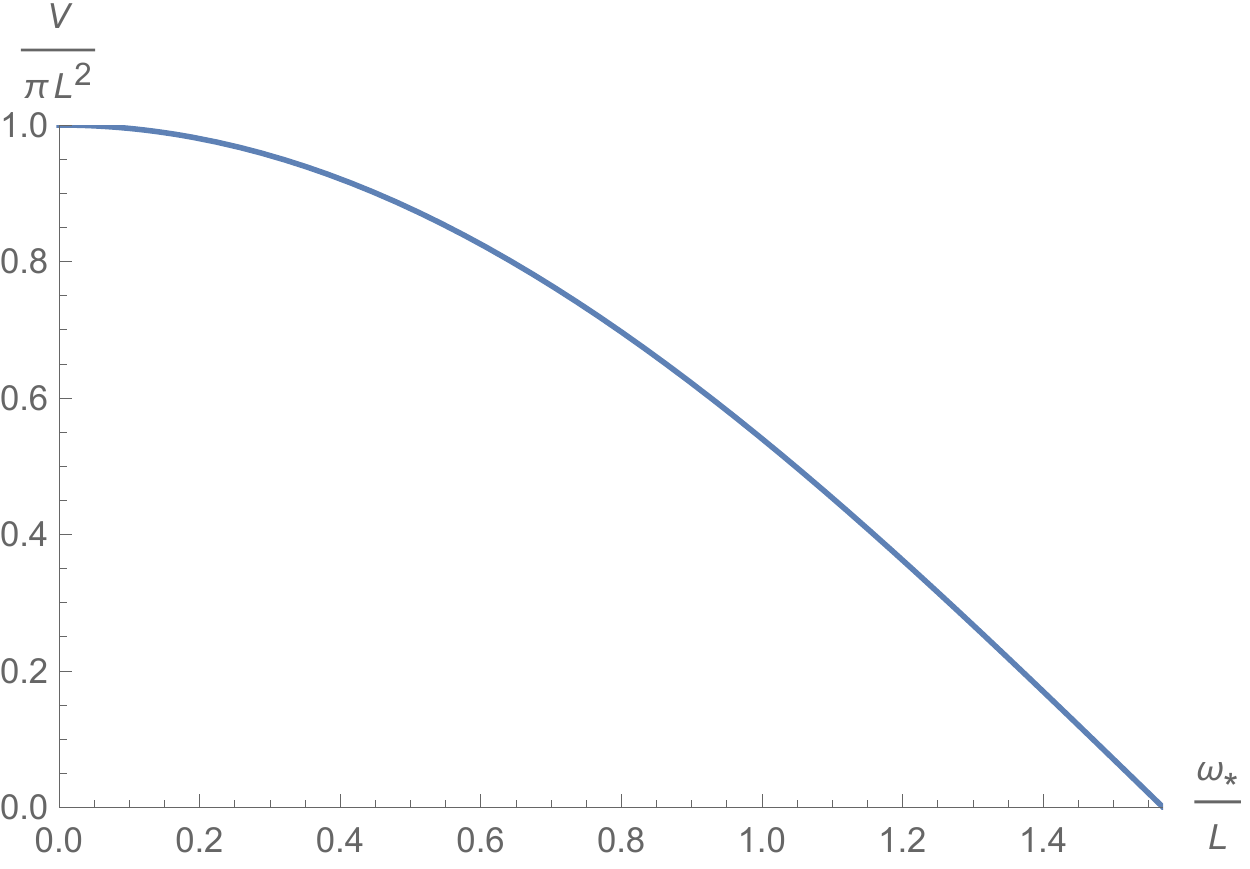}
        \caption{The volume in CV proposal, for subsystem $\mathcal{A}$ half as large as the central slice, versus the minimal bulk radial coordinates of RT surfaces.}\label{pic:CVvolume}
\end{figure}

Let's consider a field theory subsystem living on a connected region $\mathcal{A}$ in the central slice of pure dS. The spatial geometry is spherical so all the co-dimension one minimal area surfaces are half great spheres \cite{Geng:2019bnn}. As a result, if the size of $\mathcal{A}$ is not exactly equal to half of the central slice there is only one RT surface- $\mathcal{A}$ itself. In these cases, according to the CV proposal \cite{Alishahiha:2015rta} the complexity of $\mathcal{A}$ is zero if the size of $\mathcal{A}$ is smaller than half of the central slice and is the volume of the whole hemisphere if its size is larger than half of the central slice. However, if the size of $\mathcal{A}$ is precisely equal to half of the central slice there are an infinite number of RT surfaces as semicircles (with equal length) anchored at the boundary of $\mathcal{A}$. This tells us that when the size of $\mathcal{A}$ is exactly half of the central slice, we do not know how to compute the complexity using the CV proposal because we do not know which RT surface to use as the boundary (together with $\mathcal{A}$) of the bulk region whose volume computes the complexity $C(\mathcal{A})$ (see Fig.\ref{pic:CVvolume} for the volume in CV proposal as a function of the minimal bulk radial coordinate $\omega_{*}$ of the RT surface). Even though the CV complexity is not well-defined for subsystem half as large as the central slice, it is consistent with the phenomenon that if the size of $\mathcal{A}$ is growing from slightly smaller than half of the central slice to slightly bigger than, there is a jump of CV complexity from zero to the volume of the bulk hemisphere.

\subsection{Matter Deformed de-Sitter}
\label{sec:matterdeform}
In \cite{Geng:2019bnn} we did some primitive studies about the behavior of the one-parameter family of entangling surfaces when we turn on matter deformation which preserves the warped $dS_{d}$ slicing structure. However, beyond our proof that NEC tells us that the surface supposed to calculate CFT entropy has larger area than that calculating the Gibbon-Hawking entropy for the static patch, we almost know nothing about the one-parameter family of minimal surfaces with matter deformation even if it exists or not. Though, one naive thing we do know is that, with matter deformation as that in \cite{Geng:2019bnn}, the surfaces calculating CFT entropy and the static patch Gibbon-Hawking entropy are still minimal surfaces satisfying Euler-Lagrange equations for Type II and Type I minimal surfaces \cite{Geng:2019bnn} respectively.

In this section we will restrict our attention to $D=3$. We start from an observation that the area of the one-parameter family of minimal surfaces is decaying when we go deeper and deeper into the bulk if we turn on a massless scalar field. As a result, the holographic complexity in \cite{Alishahiha:2015rta} is always well-defined. 
\subsubsection{Minimal Surfaces Area Decay}
Now, for the sake of simplicity and easiness to compare, we switch to the notation in \cite{Geng:2019bnn}. For a general warped factor we have the following metric for $dS_{3}$
\begin{equation}
    ds^{2}=dr^{2}+e^{2A(r)}(-d\tau^{2}+L^{2}\cosh^{2}(\frac{\tau}{L})d\theta^{2})
\end{equation}
and as usual we focus on the time slice $\tau=0$. For Type I minimal surfaces $\theta=\theta(r)$ we have the induced metric
\begin{equation}
    ds^{2}_{\text{induced}}=dr^{2}(1+L^{2}e^{2A}\theta'(r)^{2}).
\end{equation}
Then the area reads
\begin{equation}
    Area=2\int_{r_{m}}^{r_{max}}dr\sqrt{1+L^{2}e^{2A}\theta'^{2}}
\end{equation}
which has the Euler-Lagrange equation:
\begin{equation}
    -\frac{d}{dr}\frac{L^{2}e^{2A}\theta'}{\sqrt{1+L^{2}e^{2A}\theta'^{2}}}=0
\end{equation}
with solution
\begin{equation}
    \theta'^{2}=\frac{c^2}{L^{2}e^{2A}(e^{2A}-c^{2})}
\end{equation}
for a constant $c$ parametrizing the family. The contsant $c$ can be related to the bulk maximal radial coordinate $r_{max}$ for each minimal surface through
\begin{equation}
    \theta'(r_{max})^2=\frac{c^2}{L^{2}e^{2A_{r_{max}}}(e^{2A(r_{max})}-c^{2})}=\infty.
\end{equation}
We get the expression for the area of the minimal surfaces in the family, parametrized by $r_{max}$ as
\begin{equation}
    Area(r_{max})=2\int_{0}^{r_{max}}dr\sqrt{\frac{e^{2A(r)}}{e^{2A(r)}-e^{2A(r_{max})}}}.
\end{equation}
where as in \cite{Geng:2019bnn} we take $r_{m}=0$ for the radial position of the central slice. Remember that $A'(r)<0$, as we pointed out in \cite{Geng:2019bnn}, so we can change the integration variable from $r$ to $A(r)$ using Einstein's equation:
\begin{equation}
    Area(A(r_{max}))=2\int_{A(r_{max})}^{A(0)}dA\frac{1}{\sqrt{H(r)-\frac{1}{L^{2}}+\frac{e^{-2A}}{L^{2}}}}\sqrt{\frac{e^{2A}}{e^{2A}-e^{2A(r_{max})}}}.
\end{equation}
Now we turn on massless scalar deformation and the solution of the equation of motion of the scalar field is
\begin{equation}
    \phi'(r)=Ce^{-2A}
\end{equation}
where the constant $C$ can be eliminated by the fact that at $r=0$ there is $A'=0$. Relevant equations can be found out in \cite{Geng:2019bnn}. As there, we define $\Delta=A-A(0)$, let $\Delta_{m}=A(r_{max})-A(0)$ and $x=e^{\Delta}$. We finally get
\begin{equation}
    Area(A(r_{max}))=2L\int_{e^{\Delta_{m}}}^{1}dx \frac{1}{\sqrt{x^{-4}(1-e^{-2A(0)})-1+x^{-2}e^{-2A(0)}}}\frac{1}{\sqrt{x^{2}-e^{2\Delta_{m}}}}.
\end{equation}
Before we say anything about the monotonicity of $Area(A(r_{max}))$ as a function of $r_{max}$, let's review the dependence of $\Delta_{m}$ on $r_{max}$. The warped factor $e^{A(r)}$ decreasing with $r$ (as we go from UV to IR) ends up with zero. So $\Delta_{m}$ is decreasing with $r_{max}$ and hence $e^{\Delta_{m}}\in[0,1]$.

\begin{figure}[H]
    \centering
    \includegraphics[width=12cm]{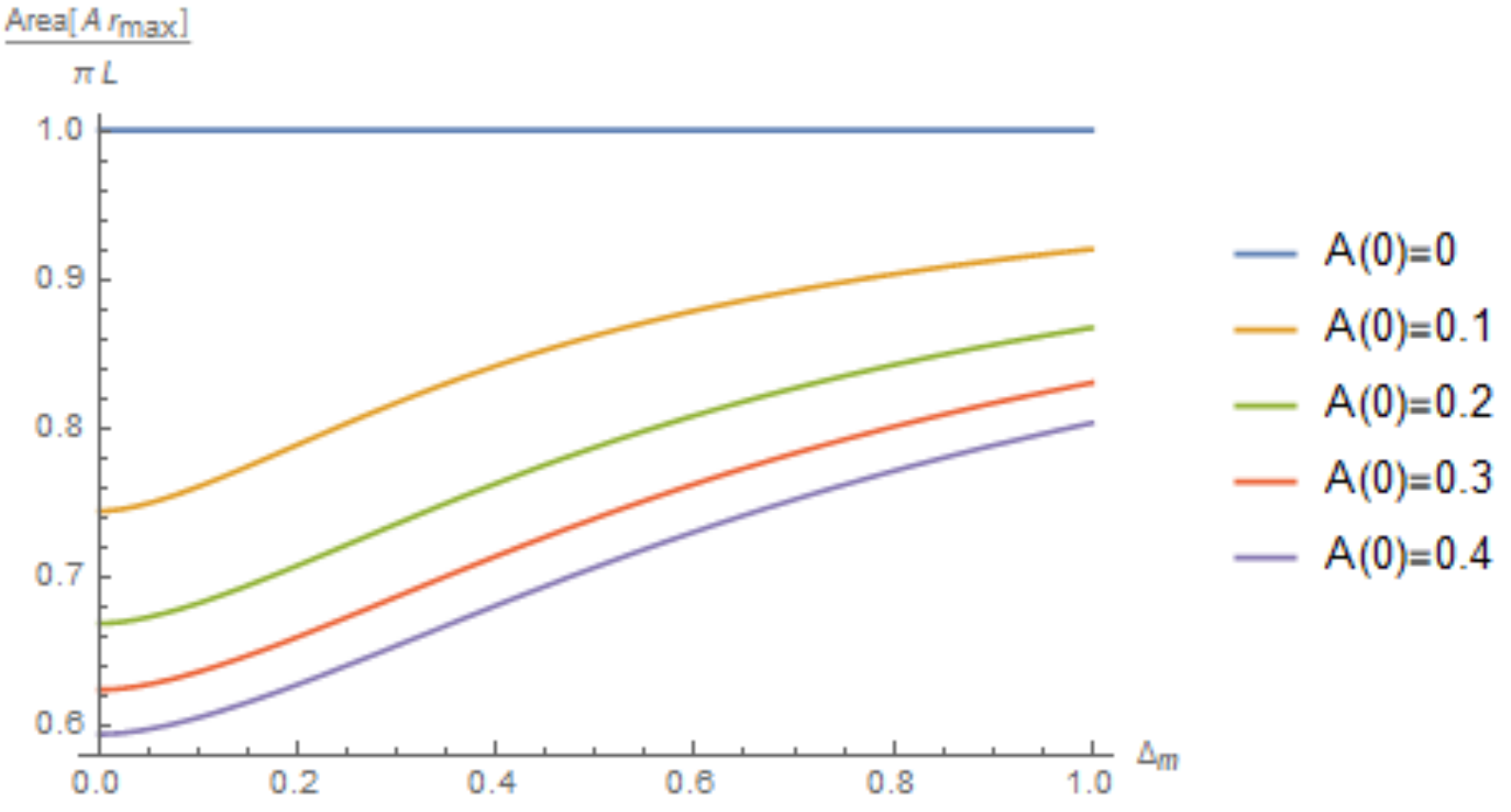}
        \caption{$\frac{Area(A(r_{max}))}{\pi L}$ versus $\Delta_{m}$. Here for the sake of presentation we only draw several curves. A(0)=0 reproduces the case for pure dS. We observe that the bigger the $A(0)s$ are the closer the curves are to each other. }\label{pic:monoto}
\end{figure}

To prove the monotonicity of $Area(A(r_{max}))$ with $r_{max}$ is not easy but numerical result shows that it is monotonically increasing with $\Delta_{m}$ (see Fig. \ref{pic:monoto}) and so decreasing with $r_{max}$. In other words, the area of the one-parameter ($r_{max}$) family of minimal surfaces decays as we go into bulk deeper and deeper from the UV cutoff (the central slice).
\subsubsection{The Implication for Complexity}
From here we see that the CV volume for the subsystem half as large as the central slice is equal to half of the volume of the bulk deformed hemisphere. Because the RT surface is now the one with minimum area which calculates the static patch Gibbon-Hawking entropy and stays at the middle of the bulk (hemisphere) anchored at the boundary of the subsystem. We might have a motivation to refine the CV proposal in a way that $C(\mathcal{A})=C(\bar{\mathcal{A}})$. But this is still not a well motivation, at least in the sense of generic states, as can be understood if we consider the case for an AdS black hole. This tells us that CV proposal has to be modified in a nontrivial way as a generic holographic dictionary.
\subsection{Cutoff de-Sitter}
\begin{figure}[H]
    \centering
    \includegraphics[width=6cm]{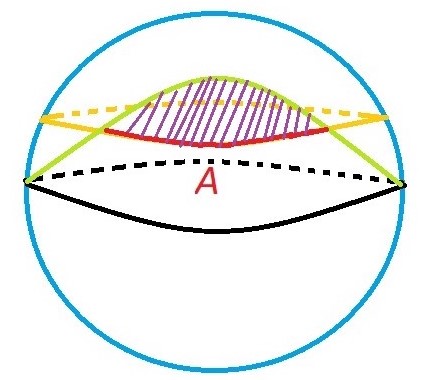}
        \caption{The cutoff slice is orange and part of it- the subsystem $\mathcal{A}$ is in red. The RT surface of $\mathcal{A}$ is the part of the green great circle over the cutoff slice. And the shaded region is the region computing complexity. }\label{pic:RTcutoff}
\end{figure}

If we consider the field theory system living on a cutoff slice, for any subsystem $\mathcal{A}$ the RT surface is different from the subsystem itself (see Fig.\ref{pic:RTcutoff}). Hence the corresponding volume in CV proposal is nonzero and so is complexity. Here the complexity changes continuously with the size of $\mathcal{A}$ which might be able to characterize some properties of $T\bar{T}+\Lambda_{2}$ deformation. Again we observe that in this case the behavior of complexity is very similar to that in pure AdS \cite{Alishahiha:2015rta}.
\subsection{Discussion}
In this section we have studied the complexity=volume proposal in \cite{Alishahiha:2004md} in the context of dS holography in several different situations. From pure dS to matter deformed dS and end up with cutoff dS.

In pure dS, we see a similar jump behavior for complexity in pure dS to that of EoP. We also notice that CV is a well-defined proposal for subsystems not exactly half as large as the central slice but it is not well-defined if the subsystem is exactly half as large as the central slice. And the problem is that there are an infinite number of RT surfaces and the regions in the bulk enclosed by each of them and the subsystem are different varying from zero to the whole bulk hemisphere. As a result, we do not know which one to use. 

Then motivated by the observation that $C(\mathcal{A})=C(\mathcal{\bar{A}})$ for equal division when we turn on a massless scalar deformation, a naive refinement of the CV proposal is still to define complexity using volume but in a way that $C(\mathcal{A})=C(\mathcal{\bar{A}})$. And so, in pure dS, the complexity of the subsystem half as large as the whole system might be equal to half of the volume of the bulk hemisphere. However, just in the sense of $C(\mathcal{A})=C(\mathcal{\bar{A}})$ this refinement for pure dS is not necessary (for example we can define that C($\mathcal{A}$)=0 if $\mathcal{A}$ is half as large as the central slice). Moreover, $C(\mathcal{A})=C(\mathcal{\bar{A}})$ is not satisfied in the case of an AdS black hole. From these considerations we claim that the CV proposal needs to be refined in a nontrivial way as a general holographic proposal. More than this, we still know very less about the behavior of the one-parameter family of minimal surfaces under matter deformation and understanding more of this would be very helpful for us to see what happens to complexity when we add matters to bulk gravity. Hence at the end complexity might be able to be used to reconstruct the bulk once we understand how to compute it from field theory (a recent progress is \cite{Geng:2019yxo}).

However, if we turn on the $T\bar{T}+\Lambda_{2}$ deformation to bring the field theory system into the bulk, the CV proposal does not seem to have any peculiarities. Therefore, the holographic CV complexity can be used to study the dual deformed field theory once we are familiar with its physical meaning on the field theory side. And the similarity of the behavior of complexity between cutoff dS and pure AdS again emphasizes the important observation in \cite{Alishahiha:2004md} that quantum gravity in dS and AdS have the same IR description.
\section{Resolving the Puzzles Using the Surface/State Correspondence}
In this section we will see that all the puzzles we discovered in the previous sections could be resolved if we treat the surface/state correspondence \cite{Miyaji:2015yva} as the fundamental principle of holography. Furthermore, we will see that we can extract some information of the entanglement structure of a state by the surface/state correspondence and this confirms our observation in \cite{Geng:2019yxo} that $T\bar{T}+\cdots$ (here in dS $\cdots$ is $\Lambda_{2}$) deformations are operating quantum circuits. These quantum circuits prepare quantum states starting from trivial states by changing their entanglement structures. We will start from a brief review of the surface/state correspondence pointing out the important ideas and conclusions we need\footnote{For more details please see the original paper \cite{Miyaji:2015yva}.}.
\subsection{Review of the Surface/State Correspondence}
The defining statement of the surface/state correspondence is that for a gravitational spacetime each codimension two convex surface is dual to a state in the Hilbert space of the system. In other words, the abstract concepts-Hilbert space and states in it are realized as the spacetime geometry.

The tricky point is the word "convex". A closed codimension two surface $\Sigma$ is convex if any codimension two extremal surface of the spacetime anchored at any codimension one submanifold of $\Sigma$ is completely inside the region enclosed by $\Sigma$. An open surface is convex if it can be embedded into a closed convex surface.

This correspondence works as follows. For a closed convex surface $\Sigma$, if it is topologically trivial (homotopic to a point) then it corresponds to a pure state $\ket{\psi(\Sigma)}$ and if it is not topologically trivial for example containing a black hole then it corresponds to a mixed state $\rho(\Sigma)$. For an open convex surface $\Sigma$ which can be embedded into a closed convex surface $\Sigma_{c}$ with complement $\tilde{\Sigma}$, it corresponds to a mixed state $\rho(\Sigma)$ which comes from $\rho(\Sigma_{c})$ by tracing out $\tilde{\Sigma}$. The extreme situation is a single point in the bulk spacetime which is argued to be dual to a state $\ket{\Omega}$ with no entanglement among its substructures\footnote{This supports the complementary picture of how $T\bar{T}$ prepares quantum states we proposed in \cite{Geng:2019yxo}}.

More than these, for a convex surface $\Sigma$ its area $A(\Sigma)$ is identified as the so called effective entropy which measures the number of effective degrees of freedom which participates in the entanglement between subsystems of $\Sigma$ and their complements on $\Sigma_{c}$ and so is smaller than the Hilbert space\footnote{This Hilbert space is not the whole Hilbert space for the whole spacetime. More details can be found at \cite{Miyaji:2015yva}.}  dimension of $\Sigma$. The entanglement entropy of $\Sigma$ with its complement $\tilde{\Sigma}$ is calculated by the minimal area surface anchored at $\partial\Sigma$. Therefore for an extremal surface all degrees of freedoms are entangled with its complement and as a result there is no entanglement among themselves. Hence, for an extremal surface $\Sigma_{A}$ the dual state is a tensor product state $\rho(\Sigma_{A})=\otimes_{i}\rho(\Sigma_{A_{i}})$ where $\Sigma_{A_{i}}$s are the small segments making up $\Sigma_{A}$, which could be well approximated by small extremal surfaces.

\subsection{The puzzles for EoP and EE}\label{sec:EESS}
The puzzle for EE is that we do not know what the entanglement entropies calculated by the RT surfaces are. Are they entanglement between the two CFTs in the DS/dS set-up? Are they entanglement among the field contents of a single CFT or a mixed concept between the two? Now using the surface/state correspondence we will abandon the field theory support of the Hilbert space and only care about if the information theoretic structure in this Hilbert space is self-consistent.

Firstly, in our pure dS context, consider the bulk central slice which is an extremal surface. The surface/state correspondence tells us that for any subsystem $\mathcal{A}$ smaller than half of the whole slice there is no entanglement among the micro-structures of it and there is only entanglement between $\mathcal{A}$ and its complement. From here by changing the size of $\mathcal{A}$  we see that for any point located on the central slice there is only entanglement between this point and its antipodal partner\footnote{This has already been noticed in \cite{Miyaji:2015yva}.}. This will be the essential point that helps us resolve all the puzzles.

This observation explains the volume law of the EE and tells that for a subsystem larger than half of the whole slice the EE is decreasing with its increasing size because then the subsystem is containing more and more entangled pairs and hence the entanglement between the subsystem and its complement is weaker and weaker. This is consistent with the RT proposal to use the smallest extremal surface to calculate EE, at least for a connected subsystem. 

Now for EoP, we still consider the situation on the right of Fig.\ref{pic:EoP}. If A and B are antipodally put on the central slice and of the same size then there is no entanglement between $A\cup B$ and its complement because $A\cup B$ contains the entangled partner of each of the point inside it. This means that the RT surface of $A\cup B$ should be empty and hence RT proposal breaks down here. The state of $A\cup B$ is a tensor product of several entangled pairs and hence a pure state. This says that $EoP$(A:B) equals to $S(A)$ and $S(B)$. General situations can be analysed similarly. Because now we are fully equipped with the entanglement structure. But the holographic definition of EoP certainly does not work here\footnote{More interestingly, the surface/state correspondence can be used to prove that the holographic definition of EoP \eqref{eq:EoP} for AdS space is the same as its information theoretic definition \eqref{eq:EoPdef}. This motivates us to believe that the surface/state correspondence can be used to find the holographic dual of any entanglement measure \cite{Horodecki:2009zz} for field theories with known gravity dual.}.
\subsection{The Puzzle for Complexity}
The puzzle for complexity is that in pure dS for a (connected) subsystem $A$ half as large as the whole central slice we do not know which RT surface to use as the boundary of the entanglement wedge whose volume computes the subregion complexity. Now we see that even with the entanglement structure fully uncovered the entanglement content of $A$ is not clear because we have to specify how precisely the size it is. If it is precisely half of the whole central slice then we are not sure if there is more than one entangled antipodal pair contained in it (which is the case for slightly bigger than half) or there is no entangled pair contained (which is the case for slightly smaller than a half) in it at all. This is consistent with the existence of an infinite number of RT surfaces for $A$. From this point of view we might have to say that the precise size of a surface are not a well defined physical concept or it is a purely classical concept which should be abandoned or operatorized in quantum gravity.
\subsection{$T\bar{T}+\cdots$ as Operating Quantum Circuits}
As we analysed in Sec.\ref{sec:EESS}, the  entanglement structure of the state dual to the DS central slice is that the "single-body" (trivial) state corresponding to a point is entangled with the "single-body" state dual to its antipodal point. However, this is clearly different from a bulk radial slice where the entanglement is not point-to-point but is smeared. The surface/state correspondence \cite{Miyaji:2015yva} tells us that the state dual to a bulk radial slice is related to the state dual to the central slice by a unitary transformation. Surprisingly, in the holographic picture of the $T\bar{T}+\Lambda_{2}$ deformation, they are related to each other by the $T\bar{T}+\Lambda_{2}$ transformation. Hence the unitary transformation suggested by the surface/state correspondence should just be the $T\bar{T}+\Lambda_{2}$ deformation. As a non-trivial self-consistent check, the $T\bar{T}+\Lambda_{2}$ transformation is reversible. This supports our observation in \cite{Geng:2019yxo} which was motivated by an analysis that CV conjecture indeed calculates the complexity of a quantum state.
\subsection{Non-locality}
For the field theory system living on the central slice, the surface/state correspondence tells us that the entanglement structure is non-local and finite. This is in contrast with the usual UV divergent behavior of entanglement entropy in local quantum field theories \cite{Witten:2018lha}. Therefore, we expect that there is some non-local features of the algebra of the field theory system. The algebra is local if for two regions $\mathcal{A}$ and $\mathcal{B}$ that the domain of dependence of $\mathcal{A}$ is contained in that of $\mathcal{B}$ then the algebra associated with $\mathcal{A}$ is contained in that with $\mathcal{B}$. A natural corollary of locality is the strong subadditivity of the entanglement entropy (more details about this point can be found in \cite{Lewkowycz:2019xse}). 

\begin{figure}[H]
    \centering
    \includegraphics[width=8cm]{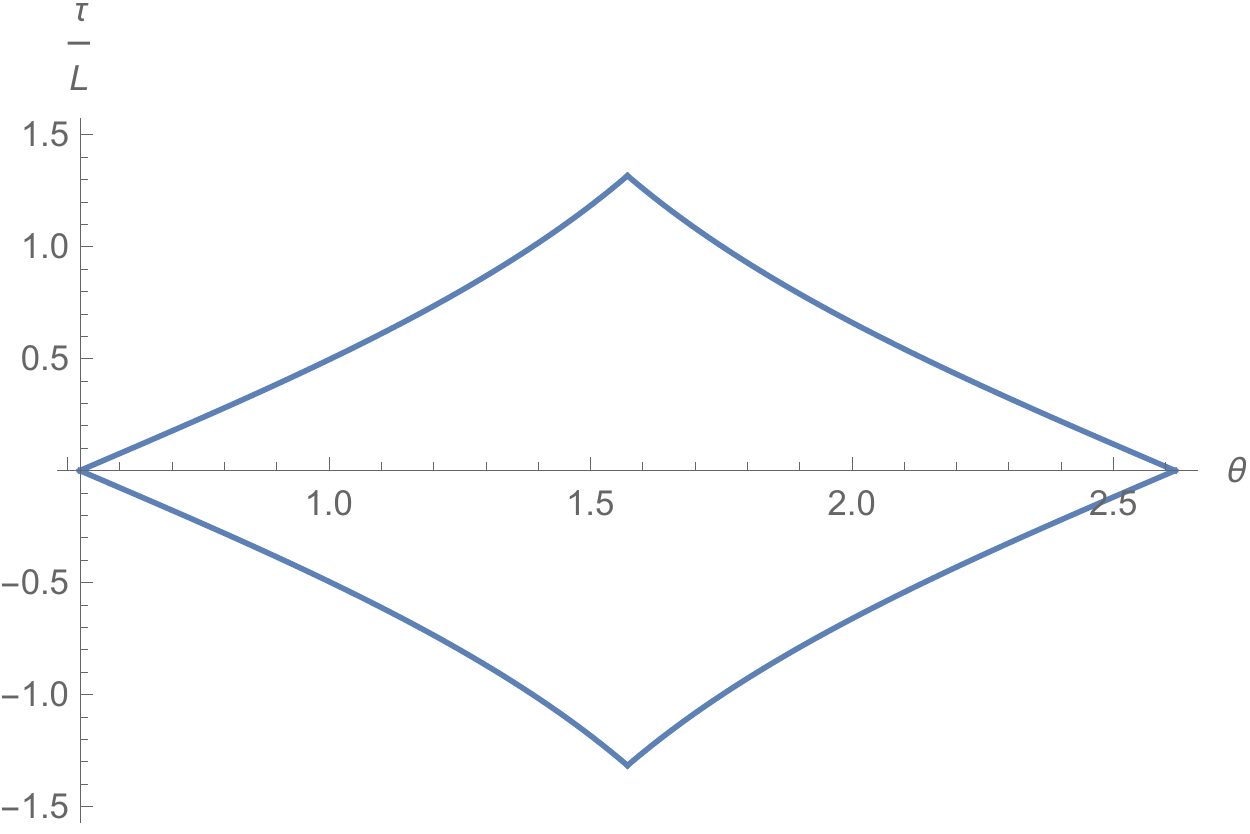}
        \caption{The domain of dependence of an interval living on the dS$_{3}$ central slice with angular size $\frac{2\pi}{3}$ and $\tau=0$. }\label{pic:domain}
\end{figure}

In this subsection we will restore the time dependence of the metric \eqref{metric} and show that the continuous version of the strong subaddivity is always satisfied and the continuous version of the ``boosted"\footnote{Our field theory system always lives on a dS$_{2}$ which has different isometry group than that of Minkowski space and hence boost should be understood as the isometry mixing space and time in a specific way in dS$_{2}$.} strong subadditivity is always violated. Before we get into details we want to show that, comparing to that in Minkowski space, for a field theory subsystem living on dS space each boundary of the causal diamond is pressed inward at the middle (See Fig.\ref{pic:domain}). 
\subsubsection{Strong Subadditivity}
Strong subadditivity can be easily proved by the monotonicity of relative entropy \cite{Witten:2018lha,Lewkowycz:2019xse} and it says that for subsystems $\mathcal{A}$ and $\mathcal{B}$ we have the following entropy inequality
\begin{equation}
    S(\mathcal{A})+S(\mathcal{B})-S(\mathcal{A}\cup\mathcal{B})-S(\mathcal{A}\cap\mathcal{B})\geq0\label{eq:ssa}.
\end{equation}
where $\mathcal{A}$ and $\mathcal{B}$ are living on the same space-like slice. Let's consider that $\mathcal{A}$ and $\mathcal{B}$ are living on a radial slice $\omega_{c}$ with $\tau=0$ then the causal diamond is
\begin{figure}[H]
    \centering
    \includegraphics[width=6cm]{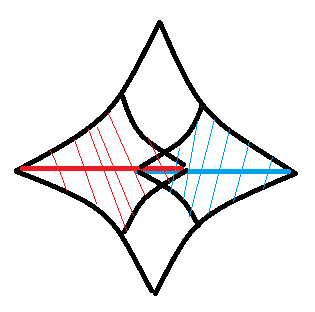}
        \caption{The causal diamond for strong subadditivity. $\mathcal{A}$ is the red interval and $\mathcal{B}$ is the blue one. The red diamond is $D[\mathcal{A}]$, the blue diamond is $D[\mathcal{B}]$, the intersection of red and blue diamonds is $D[\mathcal{A}\cap\mathcal{B}]$ and the whole diamond is $D[\mathcal{A}\cup\mathcal{B}]$.}\label{pic:causaldiamond}
\end{figure}

\begin{figure}[H]
    \centering
    \includegraphics[width=12cm]{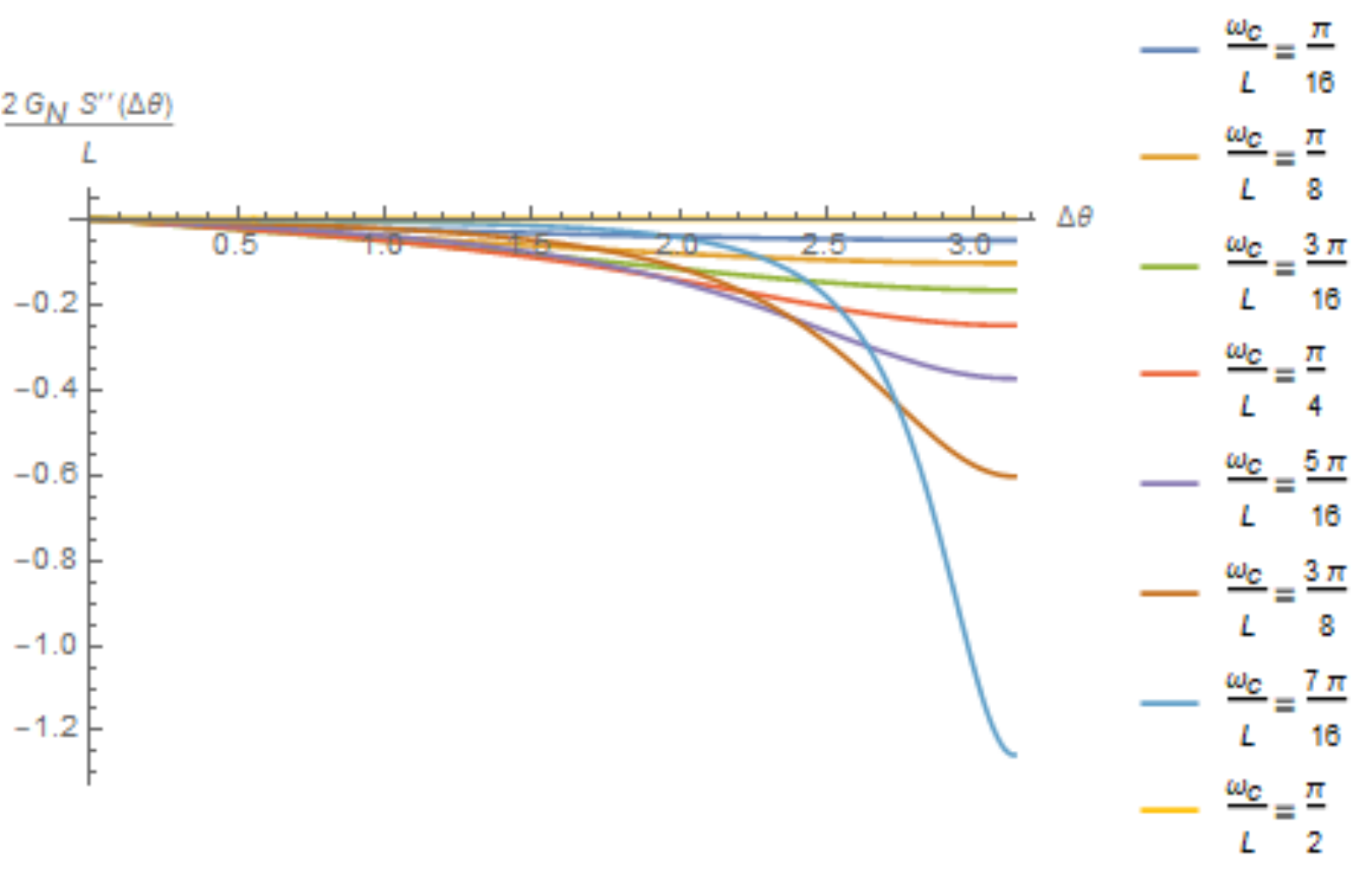}
        \caption{Strong subadditivity at different radial slices.}\label{pic:ssa}
\end{figure}

Equipped with \eqref{eq:entropysize} and taking the size of $\mathcal{A}$ and $\mathcal{B}$ to be the same, we can write \eqref{eq:ssa} as
\begin{equation}
    2S(\frac{\Delta\theta_{\mathcal{A}\cup\mathcal{B}}+\Delta\theta_{\mathcal{A}\cap\mathcal{B}}}{2},\omega_{c})-S(\Delta\theta_{\mathcal{A}\cup\mathcal{B}},\omega_{c})-S(\Delta\theta_{\mathcal{A}\cap\mathcal{B}},\omega_{c})\geq0\label{eq:ssaexp}
\end{equation}
Now taking $\mathcal{A}$ and $\mathcal{B}$ to be almost coincident with angular size $\theta$, we get the continuous version of strong subadditivity as that in \cite{Lewkowycz:2019xse}
\begin{equation}
    \partial_{\theta}^{2}S(\theta,\omega_{c})\leq0.
\end{equation}
We numerically see that this requirement is indeed satisfied (Fig.\ref{pic:ssa}). The interesting thing is that the central slice is Markov i.e. the strong subadditivity is always critically satisfied and this supports the previous observation in Sec.\ref{sec:ergodic} that the phase space of the field theory system living on the central slice is ergodic and totally random.
\subsubsection{Boosted Strong Subadditivity}
\begin{figure}[H]
    \centering
    \includegraphics[width=12cm]{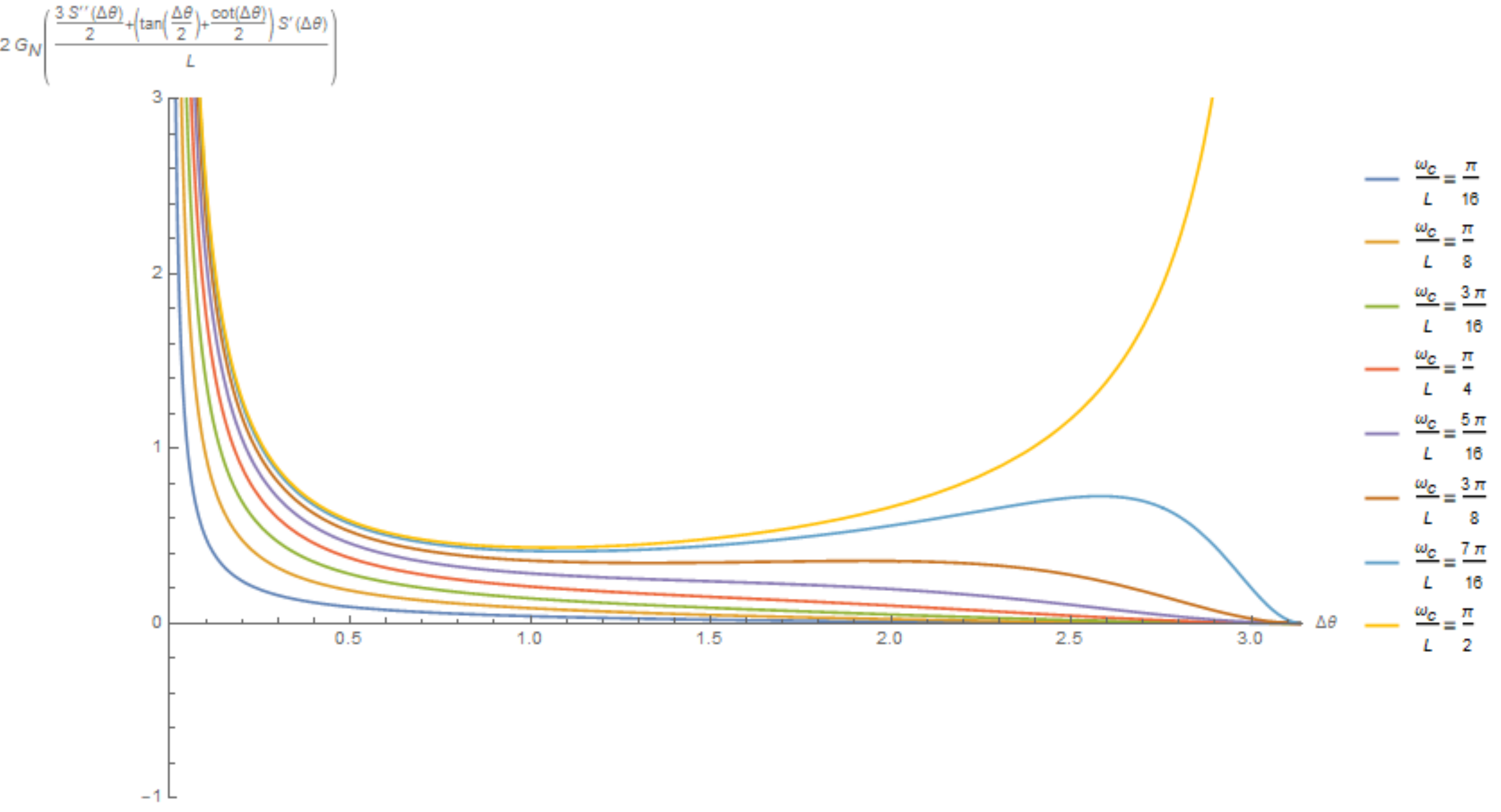}
        \caption{The boosted strong subadditivity is strongly violated for any radial slice.}\label{pic:bssa}
\end{figure}

The boosted strong subadditivity is a little bit tricky to discuss due the the fact that the field theory system is now living on dS$_{2}$. Here boost corresponds to an isometry in dS$_{2}$ which mixes space and time and preserves extremality of surfaces. This an essential observation that we are able to use formula \eqref{eq:entropysize} and the only subtle thing is to carefully identify $\Delta\theta$ for boosted intervals. Covariance tells us that we should take the boosted intervals to be minimal surfaces and use their proper lengths (without the warped-factor $\sin(\frac{\omega_{c}}{L})$) as $\Delta\theta$. This an interesting exercise and we will only show the causal diamond and results.
\begin{figure}[H]
    \centering
    \includegraphics[width=8cm]{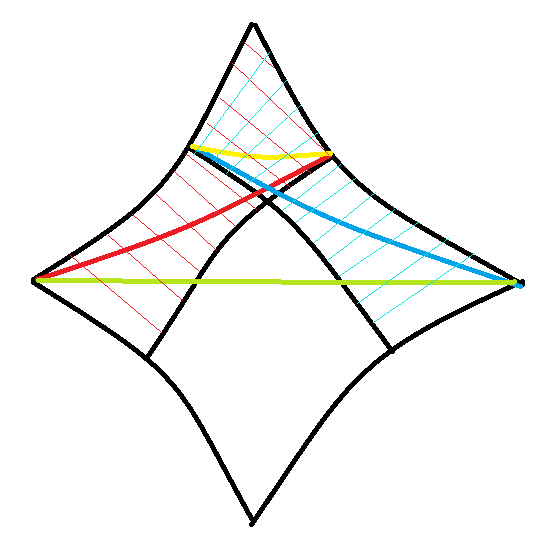}
        \caption{The causal diamond for boosted strong subadditivity. $\mathcal{A}$ is the red interval, $\mathcal{B}$ is the blue one, $\mathcal{A}\cap\mathcal{B}$ is the yellow interval and $\mathcal{A}\cup\mathcal{B}$ is the green interval. The red diamond is $D[\mathcal{A}]$, the blue diamond is $D[\mathcal{B}]$, the intersection of them is $D[\mathcal{A}\cap\mathcal{B}]$ and the whole diamond is $D[\mathcal{A}\cup\mathcal{B}]$. All intervals are minimal surfaces and the green interval is along $\tau=0$.}\label{pic:causalbssa}
\end{figure}
We take $\mathcal{A}$ and $\mathcal{B}$ to be equally large and let the $\theta$ angular size of the green interval be $\theta_{1}$ and that of the yellow interval be $\theta_{2}$\footnote{We emphasize again that this is the difference of the coordinate $\theta$ of the two end points of the interval.} then we can rewrite \eqref{eq:ssa} as
\begin{equation}
\begin{split}
    2S(\arctan&\sqrt{\cos^{2}(\frac{\theta_{1}-\theta_{2}}{2})\tan^{2}(\frac{\theta_{1}+\theta_{2}}{2})-\sin^{2}(\frac{\theta_{1}-\theta_{2}}{2})},\omega_{c})-S(\theta_{1},\omega_{c})\\&-S(2\arctan\Bigg[\tan(\frac{\theta_{2}}{2})\sqrt{\frac{1+\tan^{2}(\frac{\theta_{1}-\theta_{2}}{2})}{1-\tan^{2}\frac{\theta_{2}}{2}\tan^{2}(\frac{\theta_{1}-\theta_{2}}{2})}}\Bigg],\omega_{c})\geq0.
    \end{split}
\end{equation}
Taking the limit $\theta_{1}\rightarrow\theta_{2}=\Delta\theta$, we get the continuous version of the boosted strong subadditivity
\begin{equation}
    -(\frac{1}{2\tan(\frac{\Delta\theta}{2})}+\tan(\frac{\Delta\theta}{2}))\partial_{\Delta\theta}S(\Delta\theta,\omega_{c})-\frac{3}{2}\partial_{\Delta\theta}^{2}S(\Delta\theta,\omega_{c})\geq0.
\end{equation}
This is strongly violated (Fig.\ref{pic:bssa}). And the interesting thing is that the central slice violates the boosted strong subadditivity for all intervals smaller than or equal to half of the whole system but for other radial slices it will be saturated for the interval exactly half as large as the whole slice.
This violation of the boosted strong subadditivity reflects the non-locality of the algebra of the field theory system. The physical implication is that, in the language of Fig.\ref{pic:causalbssa}, if we boost a subsystem of the green interval to the red interval we get a larger algebra.
\section{Conclusions and Future Remarks} 
In this paper, we have studied three information theoretic quantities in the context of dS holography in terms  of the DS/dS correspondence. Several observations, calculations and motivations for future studies are provided. 

All three quantities capture the fact that quantum gravity in dS and AdS have the same near horizon (IR) description. However, they do it in different ways. EE can tell us how strong the interaction is in field theory dual and therefore the nature of the relevant degrees of freedom i.e. deconfined or confined (local or nonlocal). Also we find that from EE, in pure dS, the energy spectrum of subsystems with size smaller than half of the whole central slice behaves in an interesting way that all states are degenerate to the ground state. More than this, we find that the whole field theory system living on the central slice is likely to be chaotic or non-integrable. Therefore, we speculate the non-integrability of the field theory dual and its $T\bar{T}+\Lambda_{2}$ deformation. And it is necessary to understand this better for future studies in the low dimension version of the DS/dS correspondence. EoP and complexity perform similar behaviors in the infrared of dS and AdS. EoP can tell us how the subsystems are entangled to each other. Besides some novel behaviors of EWCS, we identify two puzzles about it if we interpret it as EoP. For complexity we find an ambiguity of using the CV proposal to compute the subregion complexity for a subregion half as large as the whole central slice in pure dS case. This problem does not exists for other situations. From here a down to earth direction is, motivated by our consideration of complexity with matter deformation, to study more about the behavior of the one-parameter family of minimal surfaces discovered in \cite{Geng:2019bnn}. For example, is the decay of the area we find in Sec.\ref{sec:matterdeform} a generic behavior in any dimension and by any physically reasonable matter deformations? A more phenomenologically oriented question is to study the implications to cosmology and low energy effective field theories in dS background from the holographic swampland bound proposed at the end of \cite{Geng:2019bnn}\footnote{Interesting progress addressing the so called swampland distance conjecture \cite{Ooguri:2018wrx} and the geometry of string theory \cite{Polchinski:1998rq,Johnson:2003gi} has recently been made in \cite{Geng:2019zsx}}. 

At the end using the surface/state correspondence we provide clear answers of all the puzzles we encountered and offer a strong support of our observation in \cite{Geng:2019yxo} about the nature of the $T\bar{T}+\cdots$ deformations. Moreover, these observations motivate us to believe that the surface/state correspondence can be used to generate the holographic dual of any correlations measure for field theories with known gravity dual. We suggest that the concept of area might have to be abandoned or operatorized in quantum gravity. Furthermore, we studied the level of non-locality as suggested by the surface/state correspondence. 

We hope that this paper will pave the way for future studies of dS holography and more generally holography of any spacetime from the information theoretic point of view\footnote{More than this, we believe that future holographic studies of dS space will tell us more interesting aspects of phenomenology and cosmology. }. Another recent and related proposal is \cite{Lewkowycz:2019xse}.
\section*{Acknowledgement}
This work is supported in part by a grant from the Simons Foundation (651440, AK). I appreciate Kristan Jensen, Andreas Karch and Eva Silverstein for useful discussions and suggestions for the draft. I am very grateful to my parents and recommenders.
\bibliographystyle{JHEP}
\bibliography{INF.bib}

\providecommand{\href}[2]{#2}\begingroup\raggedright\begin{thebibliography}{10}

\bibitem{Karch:2003em}
A.~Karch, {\it {Autolocalization in de Sitter space}},  {\em JHEP} {\bf 07}
  (2003) 050, [\href{http://arxiv.org/abs/hep-th/0305192}{{\tt
  hep-th/0305192}}].

\bibitem{Randall:1999vf}
L.~Randall and R.~Sundrum, {\it {An Alternative to compactification}},  {\em
  Phys. Rev. Lett.} {\bf 83} (1999) 4690--4693,
  [\href{http://arxiv.org/abs/hep-th/9906064}{{\tt hep-th/9906064}}].

\bibitem{Alishahiha:2004md}
M.~Alishahiha, A.~Karch, E.~Silverstein, and D.~Tong, {\it {The dS/dS
  correspondence}},  {\em AIP Conf. Proc.} {\bf 743} (2005) 393--409,
  [\href{http://arxiv.org/abs/hep-th/0407125}{{\tt hep-th/0407125}}].
  [,393(2004)].

\bibitem{Aharony:1999ti}
O.~Aharony, S.~S. Gubser, J.~M. Maldacena, H.~Ooguri, and Y.~Oz, {\it {Large N
  field theories, string theory and gravity}},  {\em Phys. Rept.} {\bf 323}
  (2000) 183--386, [\href{http://arxiv.org/abs/hep-th/9905111}{{\tt
  hep-th/9905111}}].

\bibitem{Lewkowycz:2019xse}
A.~Lewkowycz, J.~Liu, E.~Silverstein, and G.~Torroba, {\it {$T \bar T$ and EE,
  with implications for (A)dS subregion encodings}},
  \href{http://arxiv.org/abs/1909.13808}{{\tt 1909.13808}}.

\bibitem{Dong:2018cuv}
X.~Dong, E.~Silverstein, and G.~Torroba, {\it {De Sitter Holography and
  Entanglement Entropy}},  {\em JHEP} {\bf 07} (2018) 050,
  [\href{http://arxiv.org/abs/1804.08623}{{\tt 1804.08623}}].

\bibitem{Geng:2019bnn}
H.~Geng, S.~Grieninger, and A.~Karch, {\it {Entropy, Entanglement and Swampland
  Bounds in DS/dS}},  \href{http://arxiv.org/abs/1904.02170}{{\tt 1904.02170}}.

\bibitem{Gorbenko:2018oov}
V.~Gorbenko, E.~Silverstein, and G.~Torroba, {\it {dS/dS and $ T\overline{T}
  $}},  {\em JHEP} {\bf 03} (2019) 085,
  [\href{http://arxiv.org/abs/1811.07965}{{\tt 1811.07965}}].

\bibitem{Ryu:2006bv}
S.~Ryu and T.~Takayanagi, {\it {Holographic derivation of entanglement entropy
  from AdS/CFT}},  {\em Phys. Rev. Lett.} {\bf 96} (2006) 181602,
  [\href{http://arxiv.org/abs/hep-th/0603001}{{\tt hep-th/0603001}}].

\bibitem{McGough:2016lol}
L.~McGough, M.~Mezei, and H.~Verlinde, {\it {Moving the CFT into the bulk with
  $ T\overline{T} $}},  {\em JHEP} {\bf 04} (2018) 010,
  [\href{http://arxiv.org/abs/1611.03470}{{\tt 1611.03470}}].

\bibitem{Nguyen:2017yqw}
P.~Nguyen, T.~Devakul, M.~G. Halbasch, M.~P. Zaletel, and B.~Swingle, {\it
  {Entanglement of purification: from spin chains to holography}},  {\em JHEP}
  {\bf 01} (2018) 098, [\href{http://arxiv.org/abs/1709.07424}{{\tt
  1709.07424}}].

\bibitem{Takayanagi:2017knl}
T.~Takayanagi and K.~Umemoto, {\it {Entanglement of purification through
  holographic duality}},  {\em Nature Phys.} {\bf 14} (2018), no.~6 573--577,
  [\href{http://arxiv.org/abs/1708.09393}{{\tt 1708.09393}}].

\bibitem{Alishahiha:2015rta}
M.~Alishahiha, {\it {Holographic Complexity}},  {\em Phys. Rev.} {\bf D92}
  (2015), no.~12 126009, [\href{http://arxiv.org/abs/1509.06614}{{\tt
  1509.06614}}].

\bibitem{Miyaji:2015yva}
M.~Miyaji and T.~Takayanagi, {\it {Surface/State Correspondence as a
  Generalized Holography}},  {\em PTEP} {\bf 2015} (2015), no.~7 073B03,
  [\href{http://arxiv.org/abs/1503.03542}{{\tt 1503.03542}}].

\bibitem{Geng:2019yxo}
H.~Geng, {\it {$T\bar{T}$ Deformation and the Complexity=Volume Conjecture}},
  \href{http://arxiv.org/abs/1910.08082}{{\tt 1910.08082}}.

\bibitem{Hawking:1974sw}
S.~W. Hawking, {\it {Particle Creation by Black Holes}},  {\em Commun. Math.
  Phys.} {\bf 43} (1975) 199--220. [,167(1975)].

\bibitem{Bekenstein:1973ur}
J.~D. Bekenstein, {\it {Black holes and entropy}},  {\em Phys. Rev.} {\bf D7}
  (1973) 2333--2346.

\bibitem{Arias:2019zug}
C.~Arias, F.~Diaz, R.~Olea, and P.~Sundell, {\it {Liouville description of
  conical defects in dS$_4$, Gibbons-Hawking entropy as modular entropy, and
  dS$_3$ holography}},  \href{http://arxiv.org/abs/1906.05310}{{\tt
  1906.05310}}.

\bibitem{Arias:2019pzy}
C.~Arias, F.~Diaz, and P.~Sundell, {\it {De Sitter Space and Entanglement}},
  \href{http://arxiv.org/abs/1901.04554}{{\tt 1901.04554}}.

\bibitem{Grieninger:2019zts}
S.~Grieninger, {\it {Entanglement entropy and $T\bar T$ deformations beyond
  antipodal points from holography}},
  \href{http://arxiv.org/abs/1908.10372}{{\tt 1908.10372}}.

\bibitem{Harlow:2018fse}
D.~Harlow, {\it {TASI Lectures on the Emergence of Bulk Physics in AdS/CFT}},
  {\em PoS} {\bf TASI2017} (2018) 002,
  [\href{http://arxiv.org/abs/1802.01040}{{\tt 1802.01040}}].

\bibitem{Cardy:2019qao}
J.~Cardy, {\it {$T\overline T$ deformation of correlation functions}},
  \href{http://arxiv.org/abs/1907.03394}{{\tt 1907.03394}}.

\bibitem{Faulkner:2017vdd}
T.~Faulkner and A.~Lewkowycz, {\it {Bulk locality from modular flow}},  {\em
  JHEP} {\bf 07} (2017) 151, [\href{http://arxiv.org/abs/1704.05464}{{\tt
  1704.05464}}].

\bibitem{DAlessio:2016rwt}
L.~D'Alessio, Y.~Kafri, A.~Polkovnikov, and M.~Rigol, {\it {From quantum chaos
  and eigenstate thermalization to statistical mechanics and thermodynamics}},
  {\em Adv. Phys.} {\bf 65} (2016), no.~3 239--362,
  [\href{http://arxiv.org/abs/1509.06411}{{\tt 1509.06411}}].

\bibitem{2002JMP....43.4286T}
B.~M. {Terhal}, M.~{Horodecki}, D.~W. {Leung}, and D.~P. {DiVincenzo}, {\it
  {The entanglement of purification}},  {\em Journal of Mathematical Physics}
  {\bf 43} (Sep, 2002) 4286--4298,
  [\href{http://arxiv.org/abs/quant-ph/0202044}{{\tt quant-ph/0202044}}].

\bibitem{Dutta:2019gen}
S.~Dutta and T.~Faulkner, {\it {A canonical purification for the entanglement
  wedge cross-section}},  \href{http://arxiv.org/abs/1905.00577}{{\tt
  1905.00577}}.

\bibitem{Kudler-Flam:2018qjo}
J.~Kudler-Flam and S.~Ryu, {\it {Entanglement negativity and minimal
  entanglement wedge cross sections in holographic theories}},  {\em Phys.
  Rev.} {\bf D99} (2019), no.~10 106014,
  [\href{http://arxiv.org/abs/1808.00446}{{\tt 1808.00446}}].

\bibitem{Tamaoka:2018ned}
K.~Tamaoka, {\it {Entanglement Wedge Cross Section from the Dual Density
  Matrix}},  {\em Phys. Rev. Lett.} {\bf 122} (2019), no.~14 141601,
  [\href{http://arxiv.org/abs/1809.09109}{{\tt 1809.09109}}].

\bibitem{Susskind:2014rva}
L.~Susskind, {\it {Computational Complexity and Black Hole Horizons}},  {\em
  Fortsch. Phys.} {\bf 64} (2016) 44--48,
  [\href{http://arxiv.org/abs/1403.5695}{{\tt 1403.5695}}]. [Fortsch.
  Phys.64,24(2016)].

\bibitem{Horodecki:2009zz}
R.~Horodecki, P.~Horodecki, M.~Horodecki, and K.~Horodecki, {\it {Quantum
  entanglement}},  {\em Rev. Mod. Phys.} {\bf 81} (2009) 865--942,
  [\href{http://arxiv.org/abs/quant-ph/0702225}{{\tt quant-ph/0702225}}].

\bibitem{Witten:2018lha}
E.~Witten, {\it {APS Medal for Exceptional Achievement in Research: Invited
  article on entanglement properties of quantum field theory}},  {\em Rev. Mod.
  Phys.} {\bf 90} (2018), no.~4 045003,
  [\href{http://arxiv.org/abs/1803.04993}{{\tt 1803.04993}}].

\bibitem{Ooguri:2018wrx}
H.~Ooguri, E.~Palti, G.~Shiu, and C.~Vafa, {\it {Distance and de Sitter
  Conjectures on the Swampland}},  {\em Phys. Lett.} {\bf B788} (2019)
  180--184, [\href{http://arxiv.org/abs/1810.05506}{{\tt 1810.05506}}].

\bibitem{Polchinski:1998rq}
J.~Polchinski, {\em {An introduction to the bosonic string}}.
\newblock Cambridge University Press, 2007.

\bibitem{Johnson:2003gi}
C.~V. Johnson, {\em {D-branes}}.
\newblock Cambridge Monographs on Mathematical Physics. Cambridge University
  Press, 2005.

\bibitem{Geng:2019zsx}
H.~Geng, {\it {Distance Conjecture and De-Sitter Quantum Gravity}},
  \href{http://arxiv.org/abs/1910.03594}{{\tt 1910.03594}}.

\end{thebibliography}\endgroup
\end{document}